\documentclass[english,aps,aip,jcp,english,dvipsnames,preprint]{revtex4-1}
\usepackage[T1]{fontenc}
\usepackage[latin9]{inputenc}
\usepackage{float}
\usepackage{amsmath}
\usepackage{amssymb}
\usepackage{graphicx}
\usepackage{esint}
\usepackage{subscript}

\makeatletter

\providecommand{\tabularnewline}{\\}

\def\bra#1{\left<#1\right|}
\def\ket#1{\left|#1\right>}
\def\braket#1#2{\left<#1|#2\right>}

\def\part#1{\left(#1\right)}

\usepackage{amsthm}
\usepackage{graphics}
\usepackage{times}
\usepackage{bm}
\usepackage{hyperref}
\usepackage{color}
\usepackage{braket}

\setcitestyle{super}

\RequirePackage[dvipsnames]{xcolor} 

\hypersetup{
colorlinks=true,
urlcolor=Black, 
citecolor=Black,
linkcolor=Black 
}

\usepackage{braket}

\makeatother

\usepackage{babel}
\begin{document}
\title{Vibrational Investigation of Nucleobases by Means of Divide and Conquer
Semiclassical Dynamics}
\author{Fabio \surname{Gabas}}
\affiliation{Dipartimento di Chimica, Università degli Studi di Milano, via C.
Golgi 19, 20133 Milano, Italy}
\author{Giovanni \surname{Di Liberto}}
\affiliation{Dipartimento di Chimica, Università degli Studi di Milano, via C.
Golgi 19, 20133 Milano, Italy}
\author{Michele \surname{Ceotto}}
\email{michele.ceotto@unimi.it}

\affiliation{Dipartimento di Chimica, Università degli Studi di Milano, via C.
Golgi 19, 20133 Milano, Italy}
\begin{abstract}
In this work we report a computational study of the vibrational features
of four different nucleobases employing the divide-and-conquer semiclassical
initial value representation molecular dynamics method. Calculations
are performed on uracil, cytosine, thymine, and adenine. Results show
that the overall accuracy with respect to experiments is within 20
wavenumbers, regardless of the dimensionality of the nucleobase. Vibrational
estimates are accurate even in the complex case of cytosine, where
two relevant conformers are taken into account. These results are
promising in the perspective of future studies on more complex systems,
such as nucleotides or nucleobase pairs.
\end{abstract}
\maketitle

\section{Introduction\label{sec:Introduction}}

The nucleobases adenine (A), cytosine (C), guanine (G), thymine (T)
and uracil (U) represent the inner part of nucleotides, that are the
building blocks of the nucleic acids, DNA and RNA. More specifically,
a nucleotide chain constitutes a single strand that interacts with
another strand through the bases. These strands compose the secondary
and tertiary structures of the nucleic acids, as for example the famous
double helix discovered by Watson and Crick.\citep{watson1953genetical}
The interaction between nucleobases follows a specific pairing: A
and G, classified as purines, interact respectively with T (for DNA)
or U (for RNA) and C, called instead pyrimidines. The nucleobases
structure is particularly relevant since it is directly linked to
their functionality.\citep{gabelica2016nucleic} When some of them
exist in tautomeric forms, only one conformation is predominant in
nature. Also, different tautomers can bring to mispairings between
pyrimidines and purines, leading to phenomena known as mutagenesis.\citep{watson1953genetical,rein1983structure,danilov1971electronic,gabelica2016nucleic,colominas_orozco_cystosinemutations_1996,wells_wells_mutagenesis_2007}
For these reasons, an accurate investigation of the structure and
properties of all the nucleobase conformations is particularly important.
Clearly, detailed studies in condensed phase are the final goal to
understand these properties.\citep{gaigeot_sprik_IRAIMDacqueosuracil_2003,gaigeot_sprik_AIMDacqueosuracil_2004,lopez_politis_acqueosuracil2013,gustavsson_improta_uracilthymineacqueos_2006,baer_mathias_AIMDzundel_2010,marx_mathias_IRfluxional_2011,Marx_CH5plusScience_2005_understanding}
However, a preliminary investigation in vacuum is a mandatory step
in order to have a spectroscopic clear picture and to be able to discriminate
between more complex condensed phase interactions. Specifically, in
the gas phase,\citep{rijs_oomens_Nucleicbook_2015} it is possible
to study the intrinsic properties of such molecules without any intermolecular
interaction. Besides, the deep knowledge of biomolecules and their
ionic behavior in vacuum is important also for a better understanding
and designing of various spectroscopy techniques, such as mass and
vibrational spectroscopy. Thus, gas-phase vibrational spectroscopy
can certainly help in the characterization and fundamental understanding
of nucleobases.

Over the past decades a lot of experimental work has been done in
this direction, starting from the measurement of accurate spectra
for adenine, thymine and uracil,\citep{colarusso_bernath_IRnucleobasis_1997,nowak_Leszczynski_adenineIRartagged_1996,picconi_santoro_quantumclass_2013,graindourze_maes_thymineIR_1990,Les_Lapinski_thymineAr_1992,rejnek_hobza_thyminetautomers_2005,tian_xu_uraciletautomers_1999,choi_miller_uracilthymineHe_2007}
and going to the more complex spectra of cytosine and guanine, which
show more than a single relevant isomer, due to the facile tautomerism.\citep{kwiatkowski_Leszczyski_citosineIR_1996,nir_devries_guaninetautomers1_2001,choi_miller_guaninetautomers2_2006,nowak_fulara_cytosinetautomers_1989,mons_elhanine_guaninetautomerism_2002,mons_leszczyynski_guanineraretautomer_2006,marian_marian_guanineexc_2007}
These experimental spectra have been often combined with theoretical
calculations for a better interpretation. However, given the molecular
size of these systems, in most cases the theoretical prediction was
provided by harmonic frequencies calculated at the equilibrium geometry.
Even if this approach is computationally cheap and relatively immediate
to apply, it neglects all possible resonances and anharmonicities
of the potential. Recently, very detailed vibrational studies of uracil
have been presented using the canonical van Vleck second-order vibrational
perturbation theory (CVPT2), the fully automated generalized second-order
vibrational perturbation (GVPT2) approach, and the Hierarchical Intertwined
Reduced-Rank Block Power Method (HI-RRBPM), obtaining very accurate
results compared to the experiment.\citep{puzzarini2011accurate,krasnoshchekov_stefanov_CVPT2_2015,thomas_cariington_uracil_2018}
Also, Perturbation Theory (PT2) has been successful for adenine,\citep{biczysko_barone_adenine_2009}
and for the oxo isomer of citosine.\citep{walkesa_broda_citosinePT_2015}
However, at the best of our knowledge, in the case of thymine a complete
anharmonic computational spectrum is still missing.

In this work we present a vibrational spectroscopic study of four
nucleobases together with their eventual principal isomers, by means
of the semiclassical initial value representation (SCIVR) molecular
dynamics.\citep{miller_george_SCIVRelectronic_1972,Miller_PNAScomplexsystems_2005,Miller_Atom-Diatom_1970,Heller_SCnorootsearch_1991,Kay_Atomsandmolecules_2005,Tannor_book_2007,shao_makri_backfrowardnopref_1999,Pollak_Perturbationseries_2007,Pollak_Zhang_Hybridprefactor_2005,Shalashilin_Child_Coherentstates_2001,Shalashilin_Child_CCS_2004,Tao_ImportanceSampling_2014,Liu_Linearized_2015,Nandini_Church_Mixedqcl_2015,Coker_2008I2Kr}
SCIVR has been proved in recent years to be very powerful for spectroscopic
calculations.\citep{Tatchen_Pollak_Onthefly_2009,Wehrle_Vanicek_Oligothiophenes_2014,Wehrle_Vanicek_NH3_2015,Conte_Ceotto_NH3_2013,Ceotto_AspuruGuzik_Multiplecoherent_2009,Ceotto_AspuruGuzik_PCCPFirstprinciples_2009,Ceotto_Buchholz_MixedSC_2017,Ceotto_Buchholz_SAM_2018,Buchholz_Ceotto_MixedSC_2016,beguvsic_Vanicek_3thawedGaussian_2018,beguvsic_Vanicek_onthelfytimeresolvedspectra_2018,vanicek_chimiaReview_2017,wehrle_vanicek_reviewchimia_2011,Zambrano_Vanicek_Cellulardephasing_2013,Sulc_Vanicek_CellularDephasing_2012}
It does not suffer from Zero Point Energy Leakage (ZPEL)\citep{buchholz_ivanov_2018_ZPEL}
and it can be employed for vibrational eigenfunction calculations.\citep{micciarellii_ceotto_2018_eigenfunctions,micciarellii_ceotto_2019_IR,Ceotto_AspuruGuzik_Firstprinciples_2011}
In particular, the divide-and-conquer semiclassical initial value
representation (DC SCIVR) method for vibrational spectroscopy has
been employed reliably in several applications, allowing to obtain
semiclassical vibrational spectra of variously sized molecules without
any relevant loss of accuracy, and with an average deviation of 20
wavenumbers from either exact or experimental results.\citep{ceotto_conte_DCSCIVR_2017,DiLiberto_Ceotto_Jacobiano_2018}
Specifically, DC SCIVR has been applied to study challenging high-dimensional
and complex systems such as fullerene, supramolecular glycine-based
molecules and water clusters.\citep{Ceotto_watercluster_18,gabas_ceotto_glycines_2018,ceotto_conte_DCSCIVR_2017}

In the present work we calculate power spectra by performing ab-initio
molecular dynamics simulations, which has been widely and successfully
previously employed for full-dimensional on-the-fly semiclassical
applications.\citep{Gabas_Ceotto_Glycine_2017,Conte_Ceotto_NH3_2013,Ceotto_AspuruGuzik_PCCPFirstprinciples_2009,Ceotto_AspuruGuzik_Multiplecoherent_2009,gabas_ceotto_glycines_2018}
More specifically, we want to provide not only an extensive vibrational
study of nucleobases, but also to open the route to ab initio DC SCIVR
calculations on DNA-related molecules. For these goals, our quantum
mechanical vibrational estimates are compared with both experiments
and other theoretical calculations. These calculations will check
and prove DC SCIVR method feasibility and reliability, and provide
the confidence for future calculations of increasing dimensionality
up to pairs of bases and higher dimensional sequences.\citep{plutzer_vries_ATdimerSpectrum_2003,Abo_Vries_CGdimerSpectrum_2005,Nir_Vries_CGdimer_2002,fornaro_barone_uracildimer_2015}
Here, we investigate four of the five nucleobases, \textit{i.e.} uracil,
cytosine, thymine, and adenine. Since gas-phase spectra of the remaining
nucleobase, guanine, are given by a combination of its four most stable
conformers and their signals are all within few wavenumbers, guanine
is not a good benchmark for testing the accuracy of our method and
it has not been investigated.\citep{mons_elhanine_guaninetautomerism_2002,nir_DeVries_2002_guanine_properties,choi_miller_guaninetautomers2_2006}

The paper is organized as follows. Section (\ref{sec:Theory}) recalls
the DC SCIVR method for vibrational spectra calculations, together
with the computational setup. In Section (\ref{sec:Results}) we present
our results about the nucleobases and we compare them to different
experiments and other theoretical calculations. Finally, Section (\ref{sec:Conclusions})
provides some conclusions and future perspectives.

\section{Divide and Conquer Semiclassical Dynamics\label{sec:Theory}}

Vibrational power spectra are obtained via Fourier transform of the
semiclassical approximation to the survival amplitude $\braket{\chi|\chi_{t}}$

\begin{equation}
I\left(E\right)=\frac{1}{2\pi\hbar}\int dte^{\frac{i}{\hbar}Et}\braket{\chi|e^{-\frac{i}{\hbar}\hat{H}t}|\chi},\label{eq:power_spectrum}
\end{equation}
where $\left|\chi\right\rangle $ is a given reference state, $\hat{H}$
is the Hamiltonian operator and $\text{exp}\left[-i\hat{H}t/\hbar\right]$
is the quantum mechanical time-evolution operator. We calculate this
propagator using the semiclassical approximation.

Semiclassical theory takes advantage of the Feynman path integral
representation of the quantum propagator,\citep{feynman_pathintegral_1965}
in which a quantum mechanical amplitude, describing the probability
for a particle of mass $m$ to move from a certain initial state $\mathbf{q}\left(0\right)$
to a final one $\mathbf{q}\left(t\right)$ at time $t$, is calculated
as
\begin{equation}
\braket{\mathbf{q}\left(t\right)|e^{-\frac{i}{h}\widehat{H}t}|\mathbf{q}\left(0\right)}=\sqrt{\left(\frac{m}{2\pi i\hbar t}\right)^{F}}\int\wp\left[\mathbf{q}\left(t\right)\right]e^{\frac{i}{\hbar}S_{t}\left(\mathbf{q}\left(t\right),\mathbf{q}\left(0\right)\right)},\label{eq:pathintegral}
\end{equation}
where the integral of the differential $\wp\left[\mathbf{q}\left(t\right)\right]$
is given by the summation over all possible paths from $\mathbf{q}\left(0\right)$
to $\mathbf{q}\left(t\right)$. $S_{t}\left(\mathbf{q}\left(t\right),\mathbf{q}\left(0\right)\right)$
is the action of each path and $F$ is the number of degrees of freedom.
When the stationary phase approximation for the path integration is
enforced, only classical paths contribute. This approximation leads
to the well known van Vleck propagator,\citep{VanVleck_SCpropagator_1928}
where the integral is replaced by a sum over all classical paths connecting
the starting position $\mathbf{q}\left(0\right)$ to the ending one
$\mathbf{q}\left(t\right)$ in an amount of time $t$ 
\begin{equation}
\braket{\mathbf{q}\left(t\right)|e^{-\frac{i}{h}\widehat{H}t}|\mathbf{q}\left(0\right)}\sim\sqrt{\left(\frac{1}{2\pi i\hbar}\right)^{F}}\sum_{Cl\,paths}e^{\frac{i}{\hbar}S_{t}\left(\mathbf{q}\left(t\right),\mathbf{q}\left(0\right)\right)}\Biggl|\frac{\partial\mathbf{p}\left(0\right)}{\partial\mathbf{q}\left(t\right)}\Biggr|^{\frac{1}{2}}e^{-i\pi\nu/2}.
\end{equation}
 The index $\upsilon$ is called Morse, or Maslov index, and it accounts
for the sign change of the determinant $\Biggl|\frac{\partial\mathbf{p}\left(0\right)}{\partial\mathbf{q}\left(t\right)}\Biggr|$.
This version of the propagator is quite cumbersome and not practical,
since it requires to deal with a root search problem with fixed boundary
conditions. This issue has been overcome by the Initial Value Representation
trick proposed by Miller,\citep{Miller_Atom-Diatom_1970} where the
integrand of Eq.(\ref{eq:power_spectrum}) is written as
\begin{equation}
\braket{\chi|e^{-\frac{i}{h}\widehat{H}t}|\chi}\sim\sqrt{\left(\frac{1}{2\pi i\hbar}\right)^{F}}\int\int d\mathbf{q}\left(0\right)d\mathbf{p}\left(0\right)\braket{\chi|\mathbf{q}\left(t\right)}\braket{\mathbf{q}\left(0\right)|\chi}\biggl|\frac{\partial\mathbf{q}\left(t\right)}{\partial\mathbf{p}\left(0\right)}\biggr|^{\frac{1}{2}}e^{\frac{i}{\hbar}S_{t}\left(\mathbf{q}\left(0\right),\mathbf{p}\left(0\right)\right)}e^{-i\pi\nu/2}.
\end{equation}

In this version of the propagator, the survival amplitude can be obtained
by sampling trajectories with different initial conditions $\left(\mathbf{p}\left(0\right),\mathbf{q}\left(0\right)\right)$
and it is amenable to Monte Carlo integration. Later, Heller proposed
a very flexible and numerically stable representation of the semiclassical
propagator based on coherent states\citep{Heller_FrozenGaussian_1981,Heller_SCspectroscopy_1981}
of the type
\begin{equation}
\braket{\mathbf{q}|\mathbf{p}\left(t\right),\mathbf{q}\left(t\right)}=\left(\frac{\text{det}\left(\boldsymbol{\Gamma}\right)}{\pi^{F}}\right)^{\frac{1}{4}}e^{-\frac{1}{2}\left(\mathbf{q}-\mathbf{q}\left(t\right)\right)^{T}\boldsymbol{\Gamma}\left(\mathbf{q}-\mathbf{q}\left(t\right)\right)+\frac{i}{\hbar}\mathbf{p}^{T}\left(t\right)\left(\mathbf{q}-\mathbf{q}\left(t\right)\right)}\label{eq:wavepacket}
\end{equation}
where, in our simulations, the $\boldsymbol{\varGamma}$ matrix is
chosen to be constant in time and diagonal, with elements equal to
the normal mode vibrational frequencies. The wavepacket in Eq.(\ref{eq:wavepacket})
is centered at the classical position $\mathbf{q}\left(t\right)$
and momentum $\mathbf{p}\left(t\right)$, and it follows the classical
trajectory. The expression of the quantum propagator in this representation
is the well known Heller-Herman-Kluk-Kay (HHKK) propagator,\citep{Herman_Kluk_SCnonspreading_1984,Kay_SCcorrections_2006,Grossmann_Xavier_SCderivation_1998}
\begin{equation}
e^{-\frac{i}{h}\widehat{H}t}=\left(\frac{1}{2\pi\hbar}\right)^{F}\iintop d\mathbf{p}\left(0\right)d\mathbf{q}\left(0\right)C_{t}\left(\mathbf{p}\left(0\right),\mathbf{q}\left(0\right)\right)e^{\frac{i}{\hbar}S_{t}\left(\mathbf{p}\left(0\right),\mathbf{q}\left(0\right)\right)}\ket{\mathbf{p}\left(t\right),\mathbf{q}\left(t\right)}\bra{\mathbf{p}\left(0\right),\mathbf{q}\left(0\right)},\label{eq:HK_propagator}
\end{equation}
where $C_{t}\left(\mathbf{q}\left(0\right),\mathbf{p}\left(0\right)\right)$
is the pre-exponential factor and it is equal to
\begin{equation}
C_{t}\left(\mathbf{q}\left(0\right),\mathbf{p}\left(0\right)\right)=\sqrt{\mbox{det}\left[\frac{1}{2}\left(\mathbf{M}_{qq}+\boldsymbol{\Gamma}^{-1}\mathbf{M}_{pp}\boldsymbol{\Gamma}+\frac{i}{\hbar}\boldsymbol{\Gamma}^{-1}\mathbf{M}_{pq}+\frac{\hbar}{i}\mathbf{M}_{qp}\boldsymbol{\Gamma}\right)\right]}
\end{equation}
and where $\mathbf{M}_{\mathbf{ij}}$ are the monodromy (or stability)
matrix elements defined as $\mathbf{M}_{\mathbf{ij}}=\frac{\partial\mathbf{i_{t}}}{\partial\mathbf{j_{0}}}$,
where $\mathbf{i_{t}}$ is calculated at time t and $\mathbf{j_{0}}$
is calculated at time zero.\citep{Zhuang_Ceotto_Hessianapprox_2012,Ceotto_Hase_AcceleratedSC_2013}
The Monte Carlo integration in Eq. (\ref{eq:HK_propagator}) usually
needs a high number of trajectories for convergence. For systems having
more than a few degrees of freedom some filtering techniques have
been proposed to speed up the convergence.\citep{Kay_Numerical_1994,Kay_Integralexpression_1994,Kay_Multidim_1994,Wang_Miller_GeneralizedFilinov_2001}
A well established procedure is the time-averaging (TA) filtering,
in which the semiclassical integrand can be worked out to be positively
definite\citep{Kaledin_Miller_Timeaveraging_2003} by taking advantage
of the so-called separable approximation of the pre-exponential factor,
where only the phase is taken into account, i.e. $C_{t}\left(\mathbf{q}\left(0\right),\mathbf{p}\left(0\right)\right)\sim e^{\frac{i}{\hbar}\varphi_{t}}$,
and $\varphi_{t}=\text{phase\ensuremath{\left[C_{t}\left(\mathbf{q}\left(0\right),\mathbf{p}\left(0\right)\right)\right]}}$.
Using the propagator in Eq.(\ref{eq:HK_propagator}), the time-averaging
filter and the separable approximation, one obtains the following
formula for the spectral density\citep{Kaledin_Miller_Timeaveraging_2003}
\begin{equation}
I\left(E\right)=\left(\frac{1}{2\pi\hbar}\right)^{F}\iintop d\mathbf{p}\left(0\right)d\mathbf{q}\left(0\right)\frac{1}{2\pi\hbar T}\left|\intop_{0}^{T}e^{\frac{i}{\hbar}\left[S_{t}\left(\mathbf{p}\left(0\right),\mathbf{q}\left(0\right)\right)+Et+\varphi_{t}\right]}\left\langle \chi\left|\mathbf{p}\left(t\right)\mathbf{q}\left(t\right)\right.\right\rangle dt\right|^{2}.\label{eq:TA_spectrum}
\end{equation}
Eq. (\ref{eq:TA_spectrum}) has been employed to perform vibrational
estimates of small-sized molecules requiring roughly a thousand of
classical trajectories per degree of freedom to converge.\citep{Kaledin_Miller_TAmolecules_2003,Tamascelli_Ceotto_GPU_2014,Ceotto_Tantardini_Copper100_2010,DiLiberto_Ceotto_Prefactors_2016}
Unfortunately when pre-fitted Potential Energy Surfaces (PES) are
not available, the number of required trajectories is still too computational
demanding for on-the-fly or direct dynamics approaches. In this event,
only few classical trajectories can be afforded. For these reasons
in recent years, starting from Eq. (\ref{eq:TA_spectrum}), in our
group the Multiple Coherent State approach (MC SCIVR) has been developed,\citep{Ceotto_AspuruGuzik_Curseofdimensionality_2011,Ceotto_AspuruGuzik_Multiplecoherent_2009,Ceotto_AspuruGuzik_PCCPFirstprinciples_2009,Conte_Ceotto_NH3_2013}
in which by properly choosing the initial conditions of the classical
trajectories and the reference state it is possible to regain spectra
with few or even one classical trajectory, and retaining the typical
semiclassical accuracy of roughly 20 cm$^{\text{-1}}$. In the single
trajectory implementation, the initial conditions are chosen as $\left(\mathbf{p}\left(0\right),\mathbf{q}\left(0\right)\right)=\left(\mathbf{p}_{eq},\mathbf{q}_{eq}\right)$,
where $\mathbf{p}_{eq}$ and $\mathbf{q}_{eq}$ stand for the coordinates
of the reference coherent state $\left|\chi\right\rangle $. The phase
space integral of Eq.(\ref{eq:TA_spectrum}) reduces to a single trajectory
formulation for each spectroscopic peak of the type
\begin{equation}
I\left(E\right)=\left(\frac{1}{2\pi\hbar}\right)^{F}\frac{1}{2\pi\hbar T}\left|\intop_{0}^{T}e^{\frac{i}{\hbar}\left[S_{t}\left(\mathbf{p}_{eq},\mathbf{q}_{eq}\right)+Et+\varphi_{t}\left(\mathbf{p}_{eq},\mathbf{q}_{eq}\right)\right]}\left\langle \chi\left|\mathbf{p}\left(t\right),\mathbf{q}\left(t\right)\right.\right\rangle dt\right|^{2}\label{eq:MCSCIVR}
\end{equation}
where $\mathbf{q}_{eq}$ is usually chosen to be equal to the equilibrium
configuration and $\mathbf{p}_{eq}$ is set in order to provide a
total initial kinetic energy equal to the harmonic zero point energy
(ZPE), which is distributed as $p_{eq}^{i}=\sqrt{\omega_{i}}$ among
mass-scaled vibrational normal modes, where $\omega_{i}$ is the harmonic
frequency of the $i$-th mode. $\mathbf{p}\left(t\right)$ and $\mathbf{q}\left(t\right)$
are the respective time-evolved quantities. Eq.(\ref{eq:MCSCIVR})
is our working equation for MC SCIVR calculation. Moreover, in the
MC SCIVR approach, the reference state can be tailored to decompose
the spectrum mode by mode. For an $i$-th mode, we have
\begin{equation}
\ket{\chi}_{i}=\prod_{j=1}^{F}\left[\ket{p_{j,eq}^{i}\left(0\right),q_{j,eq}^{i}\left(0\right)}+\varepsilon^{i}\ket{-p_{j,eq}^{i}\left(0\right),q_{j,eq}^{i}\left(0\right)}\right],\label{eq:Multiple_Coherent_Ref_State}
\end{equation}
where the index $j$ runs over the number of vibrational degrees of
freedom F.\citep{Ceotto_AspuruGuzik_Multiplecoherent_2009,Ceotto_AspuruGuzik_Curseofdimensionality_2011}
If the $\varepsilon$ vector is set equal to one, then the zero point
energy peak is obtained (together with other even peaks), while by
setting $\varepsilon^{i}=-1$ only for a certain $i$-th mode, then
the $i$-th fundamental excitation will be enhanced, together with
the other odd overtones of $i$-th mode.\citep{Ceotto_AspuruGuzik_Curseofdimensionality_2011}
This tool is particularly helpful in presence of crowded spectra,
where peaks are very close in energy, as it commonly happens by increasing
the dimensionality of the molecule under exam. This strategy successfully
allowed to recover accurate spectra of molecules as complex as glycine.
\citep{Gabas_Ceotto_Glycine_2017}

Unfortunately, MC SCIVR runs out of steam when the system dimensionality
overcomes 25-30 degrees of freedom, because of the so-called curse
of dimensionality problem. Recently, we have addressed this issue
by proposing the DC SCIVR method,\citep{ceotto_conte_DCSCIVR_2017,DiLiberto_Ceotto_Jacobiano_2018}
where full dimensional classical simulations are projected onto sub-dimensional
spaces for semiclassical sub-dimensional spectroscopic calculations.
The same working formula of Eq.(\ref{eq:TA_spectrum}) is employed,
but formulated in terms of subspace coordinates. More specifically,
the spectral density for a M-dimensional subspace is
\begin{equation}
\widetilde{I}\left(E\right)=\left(\frac{1}{2\pi\hbar}\right)^{M}\iintop d\tilde{\mathbf{p}}\left(0\right)d\tilde{\mathbf{q}}\left(0\right)\frac{1}{2\pi\hbar T}\left|\int_{0}^{T}e^{\frac{i}{\hbar}\left[\tilde{S}_{t}\left(\tilde{\mathbf{p}}\left(0\right),\tilde{\mathbf{q}}\left(0\right)\right)+Et+\tilde{\phi}_{t}\right]}\langle\tilde{\boldsymbol{\chi}}|\tilde{\mathbf{p}}\left(t\right),\tilde{\mathbf{q}}\left(t\right)\rangle dt\right|^{2},\label{eq:DCSCIVR}
\end{equation}
where $\sim$ denotes the projected quantities. The full-dimensional
spectrum is recovered as a combination of reduced dimensionality ones.
$\braket{\mathbf{\tilde{\boldsymbol{x}}}|\tilde{\mathbf{p}}\tilde{\mathbf{q}}}=\prod_{i=1}^{M}\braket{x_{i}|p_{i}q_{i}}$
are the projected coherent states written as direct product of monodimensional
ones and involving only the degrees of freedom in the subspace. $\tilde{S}_{t}$
is the projected action, and $\tilde{\phi}_{t}$ is the phase of the
projected pre-exponential factor. This latter term is obtained from
the reduced dimensionality monodromy matrix elements upon a singular
value decomposition of the full-dimensional ones.\citep{Hinsen_Kneller_SingValueDecomp_2000,Harland_Roy_SCIVRconstrained_2003}
The projected action functional is written as
\begin{equation}
\tilde{S}_{t}\left(\tilde{\mathbf{p}}\left(0\right),\tilde{\mathbf{q}}\left(0\right)\right)=\int_{0}^{t}\left[\frac{1}{2}m\tilde{\dot{\mathbf{q}}}^{2}\left(t^{\prime}\right)-V_{S}\left(\tilde{\boldsymbol{q}}\left(t^{\prime}\right)\right)\right]dt^{\prime},
\end{equation}
where the kinetic part is trivially projected since it is naturally
separable. For the potential part, we assume that the projected potential
depend on the degrees of freedom contained in the subspace and the
remaining ones are downgraded to parameters. We include a time-dependent
external scalar field $\lambda\left(t\right)$ to ensure that the
equation of the projected potential is exact in the limit of a separable
system and it accounts for the contribution arising from the degrees
of freedom not belonging to the subspace
\begin{equation}
V_{S}\left(\tilde{\mathbf{q}}\left(t\right)\right)=V\left(\tilde{\mathbf{q}}\left(t\right);\mathbf{q}_{N_{vib}-M}^{eq}\right)+\lambda\left(t\right)
\end{equation}
\begin{equation}
\lambda\left(t\right)=V\left(\tilde{\mathbf{q}}\left(t\right);\mathbf{q}_{N_{vib}-M}\left(t\right)\right)-\left[V\left(\tilde{\mathbf{q}}\left(t\right);\mathbf{q}_{N_{vib}-M}^{eq}\right)+V\left(\mathbf{q}_{M}^{eq};\mathbf{q}_{N_{vib}-M}\left(t\right)\right)\right].
\end{equation}
Clearly, the definition of the subspaces is a critical issue within
this approach, since it is responsible for the accuracy of the action
and the monodromy matrix calculation. As for Eq.(\ref{eq:MCSCIVR}),
Eq.(\ref{eq:DCSCIVR}) can be reduced to a single trajectory formulation
and this formulation will be employed in our calculations. Next, one
desires to group in the same subspace the most interacting degrees
of freedom. We described several strategies in Ref.\citenum{DiLiberto_Ceotto_Jacobiano_2018}.
Here, we employ the one based on the Hessian matrix, since the off-diagonal
terms indicate the level of coupling between degrees of freedom. The
procedure to separate the full-dimensional space starts by defining
a time-averaged Hessian matrix $\mathbf{H}$ along a test trajectory,
where $\overline{H}_{ij}=\frac{1}{N_{steps}}\sum_{i=1}^{N_{steps}}H_{ij}$.\citep{ceotto_conte_DCSCIVR_2017,DiLiberto_Ceotto_Jacobiano_2018}
We fix a threshold parameter that is comparable with the normal mode
mass-scaled off-diagonal terms of $\overline{\boldsymbol{H}}$. If
$\overline{H}_{ij}\geq\varepsilon$, then the level of coupling is
considered significant and the degrees of freedom are enrolled in
the same subspace. Instead, if $\overline{H}_{ij}\leq\varepsilon$,
$i$ and $j$ modes are on different subspaces, unless there exists
a third mode $k$ such that $\overline{H}_{ki}\geq\varepsilon$ and/or
$\overline{H}_{kj}\geq\varepsilon$. The best possible scenario is
when there are few subspaces and they are big enough to retain most
of the couplings. However, in practice, the threshold choice is often
a trade-off between accuracy and feasibility of the semiclassical
calculation.

Molecular dynamics simulations are in this work performed using the
NWChem package\citep{Valiev_DeJong_NWChem_2010} at a DFT B3LYP/aug-cc-pvdz
level of theory,\citep{Becke_DFT_1993} which is a typical setup for
semiclassical ab-initio calculations and represents a good trade-off
between accuracy and computational overhead. For each conformer we
evolve a single trajectory for a total of 25000 atomic time units,
starting from $\left(\mathbf{p}_{eq},\mathbf{q}_{eq}\right)$, where
$\mathbf{q}_{eq}$ is the equilibrium configuration and $\mathbf{p}_{eq}$
is the momenta vector, setting each momentum to have a total initial
energy equal to the harmonic ZPE of the molecule. With the exception
of uracil molecule, we calculate the Hessian every two steps and approximate
it otherwise.\citep{Zhuang_Ceotto_Hessianapprox_2012,Ceotto_Hase_AcceleratedSC_2013}
This is a typical setup that is often enough to recover MC SCIVR and
DC SCIVR vibrational spectra with an average accuracy within 20-25
wavenumbers.\citep{Gabas_Ceotto_Glycine_2017,DiLiberto_Ceotto_Prefactors_2016,Conte_Ceotto_NH3_2013,ma_ceotto_sn2_2018}

\section{Results and Discussion\label{sec:Results}}

\begin{figure}[H]
\centering{}\includegraphics[scale=0.25]{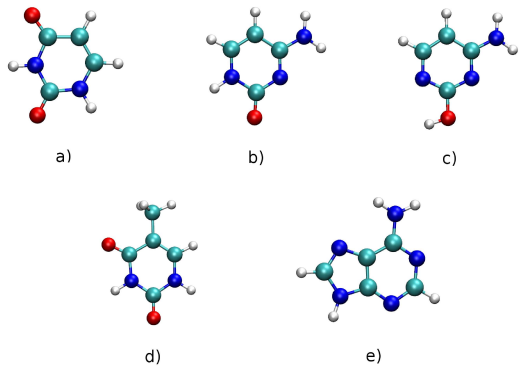}\caption{\label{fig:structure_nucleobases}Molecular structures of the simulated
nucleobases. (a) uracil, (b) and (c) respectively oxocytosine and
hydroxycytosine conformers, (d) thymine, and (e) adenine. Red: O,
grey: H, blue: N, dark green: C.}
\end{figure}

We start our investigation with the lowest dimension nucleobase, i.e.
uracil. For this system we perform both MC SCIVR and DC SCIVR calculations
to prove once more the reliability of the DC SCIVR method. Results
are compared with experiments and high level VPT2 calculations. Then,
we move to cytosine, for which there are two conformers which are
spectroscopic relevant, and to thymine and adenine vibrational spectra
and compare them to the available experimental results. With the exception
of uracil, full-dimensional MC SCIVR calculations can not be afforded
for the other nucleobases. In these cases, only DC SCIVR simulations
will be performed. The molecular structures of these molecules are
reported in Fig. \ref{fig:structure_nucleobases}.

\subsection{Uracil}

\begin{figure}[H]
\centering{}\includegraphics[scale=0.5]{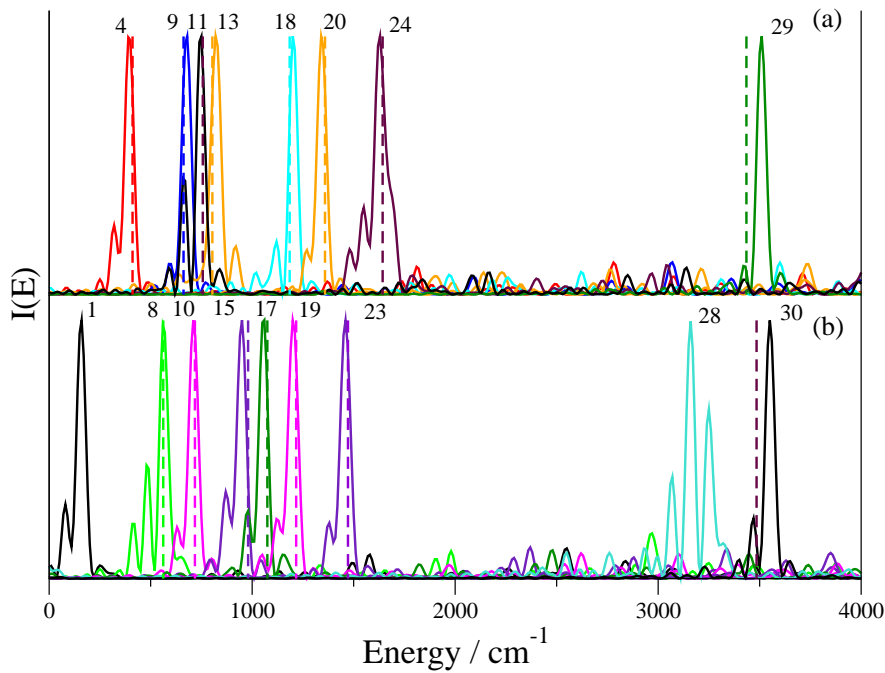}\caption{\label{fig:Uracil_15D_spectra}Panel (a) and (b): DC SCIVR fundamental
energy levels respect to the ZPE of the 17-dimensional subspace. Each
fundamental excitation was obtained by tailoring the reference state
$\ket{\chi}$ according to Eq. (\ref{eq:Multiple_Coherent_Ref_State}).
The labels of the energy levels are given according to the nomenclature
reported in Table \ref{tab:Uracil_fundamentals}. Vertical lines are
centered at experimental values.}
\end{figure}

Uracil is made of one pyrimidine ring resulting into twelve atoms.
The molecular structure of the global minimum is reported in panel
(a) of Fig. \ref{fig:structure_nucleobases}. For this molecular system
the energy difference between the global minimum (oxo form) and its
tautomer (hydroxy form) is around 45 kJ/mol.\citep{tian_xu_uraciletautomers_1999}
For this reason, we perform our semiclassical study only on the structure
reported in panel (a) of Fig. \ref{fig:structure_nucleobases}, since
it is expected to be by far the most representative one. In the past
years, this molecule was the subject of several studies, both with
experimental and theoretical approaches. Specifically, we will compare
our semiclassical vibrational estimates with experimental values\citep{graindourze_Maes_uracilederivatives_1990,szczesniak_shugar_IRuracil_1983,chin_willis_IRabinitiouracil_1984}
and GVPT2 and CVPT2 calculated energy levels.\citep{krasnoshchekov_stefanov_CVPT2_2015,puzzarini2011accurate}
As a first evaluation of the level of theory, we report in Table \ref{tab:Uracil_fundamentals}
the computed harmonic frequencies of the molecule at B3LYP/aug-cc-pvdz
level of theory together with those of Ref. \citenum{puzzarini2011accurate},
calculated using a hybrid CCSD(T)/B3LYP force field, where harmonic
CCSD(T) estimates have been corrected using GVPT2 estimates at the
level of B3LYP theory for anharmonicities. We can observe a good agreement
between these two column values, suggesting that our computational
setup is a good harmonic estimate for a possible accuracy and feasibility
of the semiclassical simulations. When moving to the semiclassical
dynamics results, each uracil fundamental frequency is evaluated by
tailoring the reference state of the semiclassical integrand according
to the MC SCIVR approach of Eq. (\ref{eq:Multiple_Coherent_Ref_State}).
Given the above mentioned dimensional limits, uracil represents also
a good benchmark for comparison between MC SCIVR and DC SCIVR. The
full-dimensional space is partitioned employing the Hessian matrix
method.\citep{ceotto_conte_DCSCIVR_2017} A threshold value equal
to $\varepsilon=4\cdot10^{-7}$ leads to a 17-dimensional subspace,
and all other subspaces are mono-dimensional. In Table \ref{tab:Uracil_fundamentals}
the computed MC SCIVR and DC SCIVR energy levels are reported. In
the fifth and sixth columns of the same Table, the GVPT2 and CVPT2
estimates are slightly more accurate than the semiclassical ones,
probably because of the higher level of ab initio theory employed.
\begin{table}[H]
\centering{}\caption{\label{tab:Uracil_fundamentals}Vibrational fundamental excitations
of uracil. Values are reported in cm$^{\text{-1}}$. The first column
reports the label of the excitation, while the second one the experimental
values from Refs.\citenum{graindourze_Maes_uracilederivatives_1990}
and \citenum{szczesniak_shugar_IRuracil_1983}. Third and fourth columns
contain our harmonic estimates and the hybrid harmonic/anharmonic
CCSD(T)/B3LYP values respectively. GVPT2 and CVPT2 estimates of Refs.
\citenum{puzzarini2011accurate} and \citenum{krasnoshchekov_stefanov_CVPT2_2015}
are reported under the VPT2 columns. The last two columns report the
energy levels calculated with MC SCIVR and DC SCIVR. ``MAE'' stands
for Mean Absolute Error.}
{\scriptsize{}}%
\begin{tabular}{ccccccccccccccccc}
{\tiny{}Label} & {\tiny{}Exp} & {\tiny{}Harmonic} & {\tiny{}Harmonic/Anharmonic} & \multicolumn{2}{c}{{\tiny{}VPT2}} & \multicolumn{2}{c}{{\tiny{}Semiclassical}} &  & {\tiny{}Label} & {\tiny{}Exp} & {\tiny{}Harmonic} & {\tiny{}Harmonic/Anharmonic} & \multicolumn{2}{c}{{\tiny{}VPT2}} & \multicolumn{2}{c}{{\tiny{}Semiclassical}}\tabularnewline
 &  & {\tiny{}B3LYP/aug-cc-pvdz} & {\tiny{}CCSD(T)/B3LYP\citep{puzzarini2011accurate}} & {\tiny{}GVPT2} & {\tiny{}CVPT2} & {\tiny{}MC SCIVR} & {\tiny{}DC SCIVR} &  &  &  & {\tiny{}B3LYP/aug-cc-pvdz} & {\tiny{}CCSD(T)/B3LYP\citep{puzzarini2011accurate}} & {\tiny{}GVPT2} & {\tiny{}CVPT2} & {\tiny{}MC SCIVR} & {\tiny{}DC SCIVR}\tabularnewline
\hline 
{\tiny{}1} &  & {\tiny{}168} & {\tiny{}140} & {\tiny{}132} & {\tiny{}139.7} & {\tiny{}156} & {\tiny{}160} &  & {\tiny{}17} & {\tiny{}1075} & {\tiny{}1083} & {\tiny{}1084} & {\tiny{}1061} & {\tiny{}1060.8} & {\tiny{}1052} & {\tiny{}1060}\tabularnewline
\hline 
{\tiny{}2} & {\tiny{}185} & {\tiny{}181} & {\tiny{}159} & {\tiny{}155} & {\tiny{}157.9} & {\tiny{}164} & {\tiny{}170} &  & {\tiny{}18} & {\tiny{}1185} & {\tiny{}1194} & {\tiny{}1205} & {\tiny{}1176} & {\tiny{}1179.9} & {\tiny{}1104} & {\tiny{}1200}\tabularnewline
\hline 
{\tiny{}3} & {\tiny{}391} & {\tiny{}395} & {\tiny{}387} & {\tiny{}386} & {\tiny{}385.6} & {\tiny{}380} & {\tiny{}370} &  & {\tiny{}19} & {\tiny{}1217} & {\tiny{}1222} & {\tiny{}1248} & {\tiny{}1221} & {\tiny{}1214.4} & {\tiny{}1208} & {\tiny{}1200}\tabularnewline
\hline 
{\tiny{}4} & {\tiny{}411} & {\tiny{}405} & {\tiny{}388} & {\tiny{}384} & {\tiny{}384.4} & {\tiny{}396} & {\tiny{}390} &  & {\tiny{}20} & {\tiny{}1359} & {\tiny{}1379} & {\tiny{}1394} & {\tiny{}1384} & {\tiny{}1382.0} & {\tiny{}1336} & {\tiny{}1340}\tabularnewline
\hline 
{\tiny{}5} & {\tiny{}516} & {\tiny{}522} & {\tiny{}517} & {\tiny{}510} & {\tiny{}510.8} & {\tiny{}524} & {\tiny{}510} &  & {\tiny{}21} & {\tiny{}1389} & {\tiny{}1402} & {\tiny{}1414} & {\tiny{}1355} & {\tiny{}1351.0} & {\tiny{}1340} & {\tiny{}1340}\tabularnewline
\hline 
{\tiny{}6} & {\tiny{}537} & {\tiny{}543} & {\tiny{}541} & {\tiny{}530} & {\tiny{}531.1} & {\tiny{}520} & {\tiny{}535} &  & {\tiny{}22} & {\tiny{}1400} & {\tiny{}1412} & {\tiny{}1427} & {\tiny{}1388} & {\tiny{}1394.1} & {\tiny{}1376} & {\tiny{}1370}\tabularnewline
\hline 
{\tiny{}7} & {\tiny{}562} & {\tiny{}560} & {\tiny{}545} & {\tiny{}549} & {\tiny{}535.3} & {\tiny{}552} & {\tiny{}550} &  & {\tiny{}23} & {\tiny{}1472} & {\tiny{}1497} & {\tiny{}1505} & {\tiny{}1466} & {\tiny{}1462.7} & {\tiny{}1464} & {\tiny{}1460}\tabularnewline
\hline 
{\tiny{}8} & {\tiny{}551} & {\tiny{}582} & {\tiny{}559} & {\tiny{}555} & {\tiny{}549.4} & {\tiny{}544} & {\tiny{}560} &  & {\tiny{}24} & {\tiny{}1643} & {\tiny{}1673} & {\tiny{}1678} & {\tiny{}1643} & {\tiny{}1642.8} & {\tiny{}1644} & {\tiny{}1630}\tabularnewline
\hline 
{\tiny{}9} & {\tiny{}662} & {\tiny{}698} & {\tiny{}670} & {\tiny{}654} & {\tiny{}651.4} & {\tiny{}676} & {\tiny{}680} &  & {\tiny{}25} & {\tiny{}1706} & {\tiny{}1757} & {\tiny{}1762} & {\tiny{}1733} & {\tiny{}1729.5} & {\tiny{}1720} & {\tiny{}1720}\tabularnewline
\hline 
{\tiny{}10} & {\tiny{}718} & {\tiny{}729} & {\tiny{}728} & {\tiny{}711} & {\tiny{}715.8} & {\tiny{}708} & {\tiny{}710} &  & {\tiny{}26} & {\tiny{}1764} & {\tiny{}1788} & {\tiny{}1790} & {\tiny{}1761} & {\tiny{}1761.2} & {\tiny{}1764} & {\tiny{}1760}\tabularnewline
\hline 
{\tiny{}11} & {\tiny{}757} & {\tiny{}764} & {\tiny{}765} & {\tiny{}746} & {\tiny{}756.1} & {\tiny{}680} & {\tiny{}750} &  & {\tiny{}27} &  & {\tiny{}3210} & {\tiny{}3218} & {\tiny{}3072} & {\tiny{}3081.0} & {\tiny{}2980} & {\tiny{}3070}\tabularnewline
\hline 
{\tiny{}12} & {\tiny{}759} & {\tiny{}771} & {\tiny{}773} & {\tiny{}751} & {\tiny{}751.8} & {\tiny{}748} & {\tiny{}750} &  & {\tiny{}28} &  & {\tiny{}3250} & {\tiny{}3253} & {\tiny{}3117} & {\tiny{}3133.6} & {\tiny{}3144} & {\tiny{}3160}\tabularnewline
\hline 
{\tiny{}13} & {\tiny{}804} & {\tiny{}821} & {\tiny{}814} & {\tiny{}793} & {\tiny{}803.2} & {\tiny{}788} & {\tiny{}820} &  & {\tiny{}29} & {\tiny{}3435} & {\tiny{}3585} & {\tiny{}3602} & {\tiny{}3436} & {\tiny{}3435.2} & {\tiny{}3480} & {\tiny{}3510}\tabularnewline
\hline 
{\tiny{}14} & {\tiny{}987} & {\tiny{}963} & {\tiny{}973} & {\tiny{}954} & {\tiny{}955.9} & {\tiny{}940} & {\tiny{}940} &  & {\tiny{}30} & {\tiny{}3485} & {\tiny{}3631} & {\tiny{}3653} & {\tiny{}3485} & {\tiny{}3486.3} & {\tiny{}3536} & {\tiny{}3550}\tabularnewline
\hline 
{\tiny{}15} & {\tiny{}958} & {\tiny{}966} & {\tiny{}968} & {\tiny{}940} & {\tiny{}947.5} & {\tiny{}952} & {\tiny{}950} &  & {\tiny{}MAE} &  & {\tiny{}26} & {\tiny{}30} & {\tiny{}10} & {\tiny{}9} & {\tiny{}22} & {\tiny{}20}\tabularnewline
\hline 
{\tiny{}16} & {\tiny{}980} & {\tiny{}988} & {\tiny{}995} & {\tiny{}978} & {\tiny{}979.9} & {\tiny{}968} & {\tiny{}970} &  &  &  &  &  &  &  &  & \tabularnewline
\hline 
\end{tabular}
\end{table}
 MC SCIVR and DC SCIVR values are very similar and close to the experimental
values as well. The MAE obtained by comparison with the experimental
results are respectively 22 and 20 wavenumbers, which is pretty much
the accuracy of other semiclassical simulations. Actually, also the
MAE for the harmonic approximation at the same level of theory is
quite similar. We think this good accuracy is accidental, because
higher level ab initio harmonic estimates (see Table 1 of Ref.\citenum{puzzarini2011accurate})
provide frequencies which are systematically higher than B3LYP ones.
Going back to the semiclassical results, we think that the DC SCIVR
MAE is slightly smaller than the more accurate MC SCIVR one, either
because of a compensation of errors or because MC SCIVR has been pushed
too close to the dimensional limit of the method. These results confirm
that the main advantage of the DC SCIVR approach is given by its portability
and applicability to higher-dimensional molecules by retaining a comparable
accuracy to that one of the MC SCIVR. \\ \indent Interestingly,
MC SCIVR is able to recover some experimental aspects which are very
hard to be reproduced, such as the exchange in energy levels between
modes 7 and 8. Namely, by employing normal mode or GVPT2 normal mode
analysis (see respectively columns three and four in Table \ref{tab:Uracil_fundamentals})
or VPT2 approaches (columns five and six of the same table), mode
7 results lower in energy than mode 8, while according to the experimental
assignment mode 7 is slightly higher in energy than mode 8. MC SCIVR
estimates agree with the experimental picture by locating mode 7 at
552 cm$^{\text{-1}}$ and mode 8 at 544 cm$^{\text{-1}}$. We believe
this peak inversion can be reproduced only by including the relevant
anharmonic effects experienced by semiclassical trajectories far from
the equilibrium configuration. \\ \indent Figure \ref{fig:Uracil_15D_spectra}
reports the spectrum of the 17-dimensional uracil subspace. Each fundamental
excitation was obtained by tailoring the reference state $\Bigl|\chi\Bigr\rangle$
of the semiclassical integrand according to Eq. (\ref{eq:Multiple_Coherent_Ref_State}).
Vertical dashed lines in the figures are centered at the available
experimental levels. The peaks are labeled according to the first
column of Table (\ref{tab:Uracil_fundamentals}) and the values are
reported in the same table.

 In summary, our results for uracil molecule show that all relevant
anharmonic effects can be reproduced and that the full dimensional
MC SCIVR calculations lead to accurate results. DC SCIVR results are
comparable and enough accurate to pursue other nucleobase investigation,
where full-dimensional MC SCIVR calculations cannot be longer afforded.

\subsection{Cytosine\label{subsec:Cytosine}}

The next nucleobase we treat is cytosine, which is constituted by
a pyrimidine ring. It differs from uracil by the presence of the amino
group in place of an oxygen atom. This molecule is made by 13 atoms,
resulting into 33 vibrational degrees of freedom. In addition to the
increased dimension, the vibrational spectroscopic investigation gets
complicated by the tautomerism between the oxocytosine and the hydroxycytosine.
Both forms are spectroscopic relevant, given the very small minimum
energy difference of few kJ/mol.\citep{kwiatkowski_Leszczyski_citosineIR_1996}
This difference is 2.1 kJ/mol at our level of DFT theory,\citep{tian_xu_uraciletautomers_1999}
and in favour of the oxo form examined in the previous section. Thus,
we perform molecular dynamics simulations of both tautomers by running
two classical trajectories, one starting from the oxocytosine isomer
and the other from the hydroxycytosine one (respectively panels (b)
and (c) of Fig. \ref{fig:structure_nucleobases}). We observe that
there is no isomeric change along the entire simulation time of 25000
au. For this nucleobase, we observe the presence of a strong coupling
between hindered rotations and other vibrational modes. For this reason,
the initial kinetic energy of the first seven lowest frequency vibrational
modes is set to zero, since rotational contributions would jeopardize
the numerical convergence of the spectra. This strategy is similar
to what has been done in previous semiclassical calculations,\citep{DiLiberto_Ceotto_Jacobiano_2018,Ceotto_watercluster_18}
and it does not represent a bias since the initial kinetic energy
of the trajectory is reduced by only 5\% below the harmonic ZPE value,
which is obviously in excess with respect to the actual ZPE. The columns
of Tables \ref{tab:Oxo_citosine_fundamentals} and \ref{tab:hydroxi_citosine_fundamentals},
labeled ``Harmonic'', report the harmonic frequencies of both isomers.
We observe from the experimental values that most of the fundamentals
of the two isomers are very similar in energy and this can be seen
already at the harmonic level. A common strategy to discriminate between
the two isomers is to look at the region around 3400-3700 wavenumbers.
In that region are present for both isomers the symmetric and asymmetric
NH\textsubscript{2} stretches, around 3500-3600 cm\textsuperscript{-1}.
Additionally the oxo form shows a peak around 3450 cm\textsuperscript{-1}
corresponding to the NH stretching, that is missing in hydroxycitosine
spectrum that instead shows a signal above 3600 cm\textsuperscript{-1},
due to the OH stretching. This trend is well reproduced by the DC
SCIVR results that can effectively discriminate between the two tautomers.
Tables \ref{tab:Oxo_citosine_fundamentals} and \ref{tab:hydroxi_citosine_fundamentals}
report the DC SCIVR computed energy levels and show the comparison
with available experimental data and other theoretical results. The
full dimensional vibrational space is divided into subspaces using
a threshold value of $\varepsilon=7\cdot10^{-7}$ for Hessian elements.
This choice generates for the oxocytosine one twenty dimensional subspace,
leaving all the others monodimensional. In the case of hydroxycytosine,
the threshold is fixed at $\varepsilon=2\cdot10^{-6}$ and the full-dimensional
space is fragmented into one eleven-dimensional, one seven-dimensional,
and all remaining monodimensional. We observe that for both isomers
the agreement with the experiment is very strict, giving a MAE around
18 cm$^{\text{-1}}$ for both systems , even if there are some frequencies
which are occasionally quite off the mark. Harmonic estimates show
a double MAE deviation.

\begin{table}[H]
\centering{}\caption{\label{tab:Oxo_citosine_fundamentals}Vibrational fundamental excitations
of the oxocytosine isomer. Values are reported in cm$^{\text{-1}}$.
The first column reports the label of the excitation, the second and
the third ones report the experimental values taken from Table 3 of
Ref.\citenum{kwiatkowski_Leszczyski_citosineIR_1996} measured in
Ne and Ar matrices respectively. Fourth column is for VPT2 energy
levels.\citep{walkesa_broda_citosinePT_2015} Fifth for Harmonic results
using a B3LYP/aug-cc-pvdz level of theory and the last column reports
the energy levels calculated with DC SCIVR at the same level of theory.
The MAEs of DC SCIVR with respect to both experiments is reported
into the last row, where the value in parenthesis is for the comparison
with the experimental values in Ar matrix.}
{\tiny{}}%
\begin{tabular}{ccccccccccccc}
{\footnotesize{}Label} & {\footnotesize{}Exp/Ne} & {\footnotesize{}Exp/Ar} & {\footnotesize{}VPT2} & {\footnotesize{}Harmonic} & {\footnotesize{}DC SCIVR} &  & {\footnotesize{}Label} & {\footnotesize{}Exp/Ne} & {\footnotesize{}Exp/Ar} & {\footnotesize{}VPT2} & {\footnotesize{}Harmonic} & {\footnotesize{}DC SCIVR}\tabularnewline
\hline 
{\footnotesize{}8} &  &  &  & {\footnotesize{}546} & {\footnotesize{}520} &  & {\footnotesize{}21} & {\footnotesize{}1237} & {\footnotesize{}1244} & {\footnotesize{}1227} & {\footnotesize{}1258} & {\footnotesize{}1220}\tabularnewline
\hline 
{\footnotesize{}9} & {\footnotesize{}571} & {\footnotesize{}575} &  & {\footnotesize{}578} & {\footnotesize{}560} &  & {\footnotesize{}22} & {\footnotesize{}1340} & {\footnotesize{}1337} & {\footnotesize{}1335} & {\footnotesize{}1351} & {\footnotesize{}1320}\tabularnewline
\hline 
{\footnotesize{}10} & {\footnotesize{}614} & {\footnotesize{}614} &  & {\footnotesize{}643} & {\footnotesize{}630} &  & {\footnotesize{}23} & {\footnotesize{}1423} & {\footnotesize{}1423} & {\footnotesize{}1408} & {\footnotesize{}1441} & {\footnotesize{}1420}\tabularnewline
\hline 
{\footnotesize{}11} & {\footnotesize{}717} & {\footnotesize{}716} &  & {\footnotesize{}726} & {\footnotesize{}710} &  & {\footnotesize{}24} & {\footnotesize{}1475} & {\footnotesize{}1475} & {\footnotesize{}1472} & {\footnotesize{}1497} & {\footnotesize{}1480}\tabularnewline
\hline 
{\footnotesize{}12} & {\footnotesize{}749} & {\footnotesize{}747} &  & {\footnotesize{}768} & {\footnotesize{}750} &  & {\footnotesize{}25} & {\footnotesize{}1540} & {\footnotesize{}1539} & {\footnotesize{}1521} & {\footnotesize{}1566} & {\footnotesize{}1550}\tabularnewline
\hline 
{\footnotesize{}13} & {\footnotesize{}767} & {\footnotesize{}747} &  & {\footnotesize{}771} & {\footnotesize{}760} &  & {\footnotesize{}26} & {\footnotesize{}1602} & {\footnotesize{}1598} & {\footnotesize{}1588} & {\footnotesize{}1629} & {\footnotesize{}1590}\tabularnewline
\hline 
{\footnotesize{}14} & {\footnotesize{}784} & {\footnotesize{}818} &  & {\footnotesize{}786} & {\footnotesize{}780} &  & {\footnotesize{}27} & {\footnotesize{}1659} & {\footnotesize{}1656} & {\footnotesize{}1647} & {\footnotesize{}1685} & {\footnotesize{}1660}\tabularnewline
\hline 
{\footnotesize{}15} &  &  &  & {\footnotesize{}921} & {\footnotesize{}890} &  & {\footnotesize{}28} & {\footnotesize{}1730} & {\footnotesize{}1733} & {\footnotesize{}1736} & {\footnotesize{}1755} & {\footnotesize{}1750}\tabularnewline
\hline 
{\footnotesize{}16} &  &  &  & {\footnotesize{}964} & {\footnotesize{}940} &  & {\footnotesize{}29} &  &  & {\footnotesize{}3037} & {\footnotesize{}3202} & {\footnotesize{}3140}\tabularnewline
\hline 
{\footnotesize{}17} &  &  &  & {\footnotesize{}984} & {\footnotesize{}980} &  & {\footnotesize{}30} &  &  & {\footnotesize{}3092} & {\footnotesize{}3227} & {\footnotesize{}3120}\tabularnewline
\hline 
{\footnotesize{}18} &  & {\footnotesize{}1088} &  & {\footnotesize{}1082} & {\footnotesize{}1060} &  & {\footnotesize{}31} & {\footnotesize{}3457} & {\footnotesize{}3441} & {\footnotesize{}3460} & {\footnotesize{}3596} & {\footnotesize{}3510}\tabularnewline
\hline 
{\footnotesize{}19} & {\footnotesize{}1103} & {\footnotesize{}1090} & {\footnotesize{}1106} & {\footnotesize{}1120} & {\footnotesize{}1110} &  & {\footnotesize{}32} & {\footnotesize{}3474} & {\footnotesize{}3472} & {\footnotesize{}3477} & {\footnotesize{}3611} & {\footnotesize{}3510}\tabularnewline
\hline 
{\footnotesize{}20} & {\footnotesize{}1198} & {\footnotesize{}1196} & {\footnotesize{}1193} & {\footnotesize{}1208} & {\footnotesize{}1190} &  & {\footnotesize{}33} & {\footnotesize{}3575} & {\footnotesize{}3564} & {\footnotesize{}3610} & {\footnotesize{}3741} & {\footnotesize{}3580}\tabularnewline
\hline 
 &  &  &  &  &  &  & {\footnotesize{}MAE} &  &  & {\footnotesize{}10(13)} & {\footnotesize{}38(41)} & {\footnotesize{}13(18)}\tabularnewline
\hline 
\end{tabular}
\end{table}
\begin{table}[H]
\centering{}\caption{\label{tab:hydroxi_citosine_fundamentals} The same as Table \ref{tab:Oxo_citosine_fundamentals}
but for the hydroxycytosine isomer.}
{\tiny{}}%
\begin{tabular}{ccccccccccc}
{\scriptsize{}Label} & {\scriptsize{}Exp/Ne} & {\scriptsize{}Exp/Ar} & {\scriptsize{}Harmonic} & {\scriptsize{}DC SCIVR} &  & {\scriptsize{}Label} & {\scriptsize{}Exp/Ne} & {\scriptsize{}Exp/Ar} & {\scriptsize{}Harmonic} & {\scriptsize{}DC SCIVR}\tabularnewline
\hline 
{\scriptsize{}8} & {\scriptsize{}553} & {\scriptsize{}557} & {\scriptsize{}560} & {\scriptsize{}540} &  & {\scriptsize{}21} & {\scriptsize{}1258} & {\scriptsize{}1257} & {\scriptsize{}1298} & {\scriptsize{}1270}\tabularnewline
\hline 
{\scriptsize{}9} & {\scriptsize{}525} & {\scriptsize{}520} & {\scriptsize{}563} & {\scriptsize{}540} &  & {\scriptsize{}22} & {\scriptsize{}1338} & {\scriptsize{}1333} & {\scriptsize{}1342} & {\scriptsize{}1330}\tabularnewline
\hline 
{\scriptsize{}10} & {\scriptsize{}600} & {\scriptsize{}601} & {\scriptsize{}602} & {\scriptsize{}570} &  & {\scriptsize{}23} & {\scriptsize{}1382} & {\scriptsize{}1379} & {\scriptsize{}1399} & {\scriptsize{}1380}\tabularnewline
\hline 
{\scriptsize{}11} & {\scriptsize{}711} & {\scriptsize{}710} & {\scriptsize{}721} & {\scriptsize{}730} &  & {\scriptsize{}24} & {\scriptsize{}1441} & {\scriptsize{}1439} & {\scriptsize{}1459} & {\scriptsize{}1440}\tabularnewline
\hline 
{\scriptsize{}12} & {\scriptsize{}784} & {\scriptsize{}751} & {\scriptsize{}790} & {\scriptsize{}790} &  & {\scriptsize{}25} & {\scriptsize{}1495} & {\scriptsize{}1496} & {\scriptsize{}1517} & {\scriptsize{}1490}\tabularnewline
\hline 
{\scriptsize{}13} & {\scriptsize{}796} & {\scriptsize{}781} & {\scriptsize{}794} & {\scriptsize{}770} &  & {\scriptsize{}26} & {\scriptsize{}1576} & {\scriptsize{}1575} & {\scriptsize{}1606} & {\scriptsize{}1570}\tabularnewline
\hline 
{\scriptsize{}14} & {\scriptsize{}809} & {\scriptsize{}807} & {\scriptsize{}818} & {\scriptsize{}830} &  & {\scriptsize{}27} & {\scriptsize{}1592} & {\scriptsize{}1589} & {\scriptsize{}1630} & {\scriptsize{}1610}\tabularnewline
\hline 
{\scriptsize{}15} & {\scriptsize{}948} & {\scriptsize{}955} & {\scriptsize{}986} & {\scriptsize{}970} &  & {\scriptsize{}28} & {\scriptsize{}1625} & {\scriptsize{}1622} & {\scriptsize{}1656} & {\scriptsize{}1640}\tabularnewline
\hline 
{\scriptsize{}16} & {\scriptsize{}948} & {\scriptsize{}955} & {\scriptsize{}993} & {\scriptsize{}970} &  & {\scriptsize{}29} &  &  & {\scriptsize{}3165} & {\scriptsize{}3080}\tabularnewline
\hline 
{\scriptsize{}17} & {\scriptsize{}989} & {\scriptsize{}980} & {\scriptsize{}997} & {\scriptsize{}980} &  & {\scriptsize{}30} &  &  & {\scriptsize{}3207} & {\scriptsize{}3090}\tabularnewline
\hline 
{\scriptsize{}18} & {\scriptsize{}1085} & {\scriptsize{}1083} & {\scriptsize{}1092} & {\scriptsize{}1090} &  & {\scriptsize{}31} & {\scriptsize{}3461} & {\scriptsize{}3446} & {\scriptsize{}3596} & {\scriptsize{}3480}\tabularnewline
\hline 
{\scriptsize{}19} & {\scriptsize{}1113} & {\scriptsize{}1110} & {\scriptsize{}1120} & {\scriptsize{}1100} &  & {\scriptsize{}32} & {\scriptsize{}3575} & {\scriptsize{}3564} & {\scriptsize{}3736} & {\scriptsize{}3600}\tabularnewline
\hline 
{\scriptsize{}20} & {\scriptsize{}1198} & {\scriptsize{}1196} & {\scriptsize{}1241} & {\scriptsize{}1220} &  & {\scriptsize{}33} & {\scriptsize{}3618} & {\scriptsize{}3592} & {\scriptsize{}3770} & {\scriptsize{}3670}\tabularnewline
\hline 
 &  &  &  &  &  & {\scriptsize{}MAE} &  &  & {\scriptsize{}41(36)} & {\scriptsize{}16(18)}\tabularnewline
\hline 
\end{tabular}
\end{table}

\begin{figure}[H]
\centering{}\includegraphics[scale=0.5]{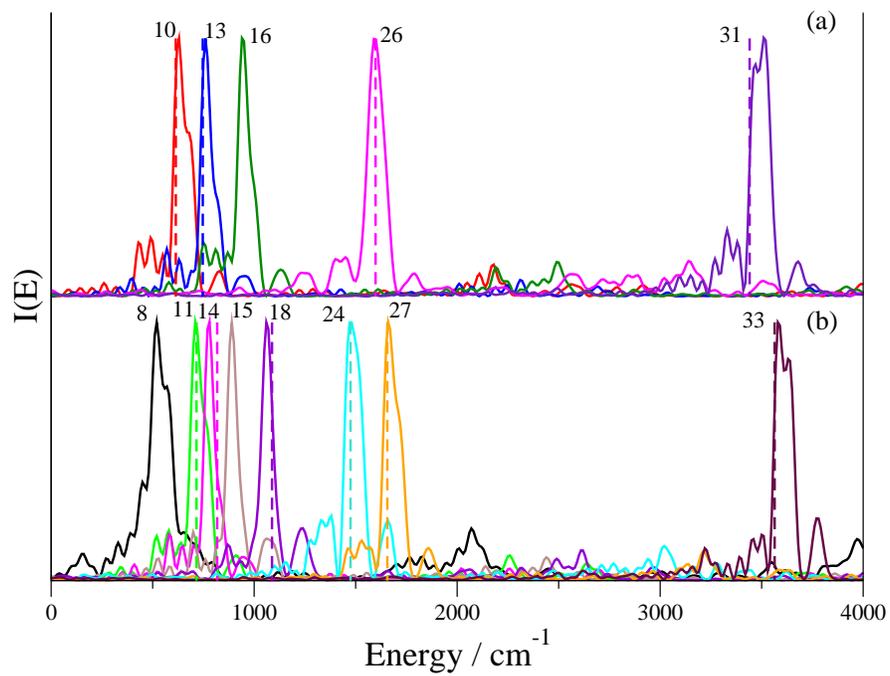}\caption{\label{fig:oxo_citosine_XD_3}Panel (a) and (b): The same as in Fig.
\ref{fig:Uracil_15D_spectra} but for the 20-dimensional subspace
of the oxocytosine isomer. Labels of the energy levels follow the
nomenclature reported in Table \ref{tab:Oxo_citosine_fundamentals}.}
\end{figure}

\begin{figure}[H]
\centering{}\includegraphics[scale=0.5]{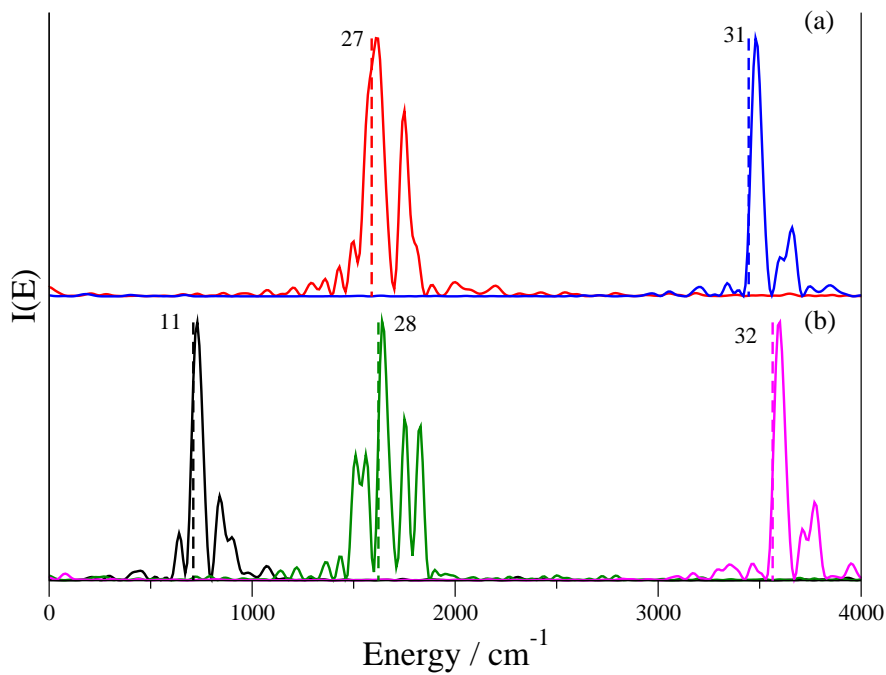}\caption{\label{fig:hysroxi_citosine_XD_3}Panel (a) and (b): The same as Fig.
\ref{fig:oxo_citosine_XD_3} but for the 11-dimensional hydroxycytosine
isomer subspaces. Missing peaks in the figure are those with initial
momenta set to 0. Labels of the energy levels follow the nomenclature
reported in Table \ref{tab:hydroxi_citosine_fundamentals}.}
\end{figure}

Figure \ref{fig:oxo_citosine_XD_3} shows DC SCIVR spectra for a 20-dimensional
subspace of the oxocytosine isomer, while Fig. \ref{fig:hysroxi_citosine_XD_3}
reports the computed spectra for a 11-dimensional subspaces of the
hydroxycytosine isomer. We observe that the number of peaks in Fig.s
\ref{fig:oxo_citosine_XD_3} and \ref{fig:hysroxi_citosine_XD_3}
is less than the subspace-dimension, since such subspace contains
one low index mode (1-7). These figures show neat spectroscopic signals
with several overtones reproduced, especially Fig. \ref{fig:hysroxi_citosine_XD_3}.
So far we have looked at the lowest dimensional couple of nucleobases
made by a pyrimidine ring. Our results agree with experiments with
an average deviation of 15-20 wavenumbers, in line with previous semiclassical
simulations, and are comparable with other theoretical methods. In
the next section we look at thymine, the last pyrimidine-based nucleobase.

\subsection{Thymine}

Thymine is the highest dimension pyrimidine nucleobase. The structure
is similar to uracil, where one hydrogen atom is substituted by a
methyl group. Because of their resemblance, thymine and uracil show
some similarities in biological processes and both link to adenine.
The molecule is made of 15 atoms leading to 39 vibrational degrees
of freedom. We calculate the vibrational spectra using classical trajectories
starting from the global minimum structure, as we have done for uracil
in Section \ref{subsec:Cytosine}, since the hydroxy tautomer minimum
is about 44 kJ/mol less stable,\citep{rejnek_hobza_thyminetautomers_2005}
a value similar to the one between uracil tautomers (45 kJ/mol).\citep{tian_xu_uraciletautomers_1999}
This is expected given their structural similarity. The geometry of
the molecule at equilibrium configuration is reported in Panel (d)
of Fig. \ref{fig:structure_nucleobases}. The Hessian partitioning
method leads to a 18-dimensional subspace, a seven-dimensional one
and all other are mono-dimensional, when employing a threshold parameter
$\varepsilon=8\cdot10^{-7}$. DC SCIVR frequencies are calculated
for each subspace. 
\begin{table}[H]
\caption{\label{tab:timine_fundamentals} Two sides table for the vibrational
frequencies of thymine. First column reports the label of the excitation.
Columns two and three are for gas phase spectra of the isolated\citep{colarusso_bernath_IRnucleobasis_1997}
and Ar-tagged thymine.\citep{Les_Lapinski_thymineAr_1992,graindourze_maes_thymineIR_1990}
Column four reports the harmonic estimate at the level of B3LYP/aug-cc-pvdz
and the last column our computed DC SCIVR energy levels at the same
level of theory. MAE is reported on the last row in comparison with
both series of experiments. Values are reported in cm$^{\text{-1}}$.}

\centering{}{\scriptsize{}}%
\begin{tabular}{ccccccccccc}
{\scriptsize{}mode} & {\scriptsize{}Gas} & {\scriptsize{}Gas/Ar} & {\scriptsize{}Harmonic} & {\scriptsize{}DC SCIVR} &  & {\scriptsize{}mode} & {\scriptsize{}Gas} & {\scriptsize{}Gas/Ar} & {\scriptsize{}Harmonic} & {\scriptsize{}DC SCIVR}\tabularnewline
\hline 
{\scriptsize{}1} &  &  & {\scriptsize{}130} & {\scriptsize{}110} &  & {\scriptsize{}21} & {\scriptsize{}1078} & {\scriptsize{}1087} & {\scriptsize{}1153} & {\scriptsize{}1120}\tabularnewline
{\scriptsize{}2} &  &  & {\scriptsize{}162} & {\scriptsize{}150} &  & {\scriptsize{}22} & {\scriptsize{}1178} & {\scriptsize{}1184} & {\scriptsize{}1196} & {\scriptsize{}1170}\tabularnewline
\hline 
{\scriptsize{}3} &  &  & {\scriptsize{}168} & {\scriptsize{}150} &  & {\scriptsize{}23} &  & {\scriptsize{}1221} & {\scriptsize{}1232} & {\scriptsize{}1190}\tabularnewline
\hline 
{\scriptsize{}4} &  & {\scriptsize{}280} & {\scriptsize{}281} & {\scriptsize{}290} &  & {\scriptsize{}24} &  & {\scriptsize{}1357} & {\scriptsize{}1367} & {\scriptsize{}1380}\tabularnewline
\hline 
{\scriptsize{}5} &  &  & {\scriptsize{}302} & {\scriptsize{}310} &  & {\scriptsize{}25} &  & {\scriptsize{}1395} & {\scriptsize{}1397} & {\scriptsize{}1360}\tabularnewline
\hline 
{\scriptsize{}6} &  & {\scriptsize{}391} & {\scriptsize{}394} & {\scriptsize{}380} &  & {\scriptsize{}26} & {\scriptsize{}1393} & {\scriptsize{}1389} & {\scriptsize{}1408} & {\scriptsize{}1390}\tabularnewline
\hline 
{\scriptsize{}7} &  & {\scriptsize{}394} & {\scriptsize{}407} & {\scriptsize{}390} &  & {\scriptsize{}27} & {\scriptsize{}1409} & {\scriptsize{}1405} & {\scriptsize{}1419} & {\scriptsize{}1380}\tabularnewline
\hline 
{\scriptsize{}8} & {\scriptsize{}462} & {\scriptsize{}455} & {\scriptsize{}461} & {\scriptsize{}460} &  & {\scriptsize{}28} &  & {\scriptsize{}1437} & {\scriptsize{}1448} & {\scriptsize{}1430}\tabularnewline
\hline 
{\scriptsize{}9} &  & {\scriptsize{}540} & {\scriptsize{}546} & {\scriptsize{}530} &  & {\scriptsize{}29} &  & {\scriptsize{}1451} & {\scriptsize{}1473} & {\scriptsize{}1460}\tabularnewline
\hline 
{\scriptsize{}10} & {\scriptsize{}541} & {\scriptsize{}545} & {\scriptsize{}583} & {\scriptsize{}570} &  & {\scriptsize{}30} & {\scriptsize{}1463} & {\scriptsize{}1472} & {\scriptsize{}1497} & {\scriptsize{}1460}\tabularnewline
\hline 
{\scriptsize{}11} &  &  & {\scriptsize{}603} & {\scriptsize{}600} &  & {\scriptsize{}31} & {\scriptsize{}1668} & {\scriptsize{}1684} & {\scriptsize{}1700} & {\scriptsize{}1690}\tabularnewline
\hline 
{\scriptsize{}12} & {\scriptsize{}689} & {\scriptsize{}662} & {\scriptsize{}700} & {\scriptsize{}670} &  & {\scriptsize{}32} & {\scriptsize{}1725} & {\scriptsize{}1712} & {\scriptsize{}1741} & {\scriptsize{}1750}\tabularnewline
\hline 
{\scriptsize{}13} &  & {\scriptsize{}727} & {\scriptsize{}735} & {\scriptsize{}710} &  & {\scriptsize{}33} & {\scriptsize{}1772} & {\scriptsize{}1768} & {\scriptsize{}1784} & {\scriptsize{}1760}\tabularnewline
\hline 
{\scriptsize{}14} & {\scriptsize{}755} & {\scriptsize{}754} & {\scriptsize{}757} & {\scriptsize{}750} &  & {\scriptsize{}34} &  & {\scriptsize{}2855} & {\scriptsize{}3038} & {\scriptsize{}2955}\tabularnewline
\hline 
{\scriptsize{}15} & {\scriptsize{}767} & {\scriptsize{}764} & {\scriptsize{}779} & {\scriptsize{}770} &  & {\scriptsize{}35} & {\scriptsize{}2941} & {\scriptsize{}2940} & {\scriptsize{}3097} & {\scriptsize{}2970}\tabularnewline
\hline 
{\scriptsize{}16} & {\scriptsize{}804} & {\scriptsize{}800} & {\scriptsize{}802} & {\scriptsize{}780} &  & {\scriptsize{}36} & {\scriptsize{}2984} & {\scriptsize{}2971} & {\scriptsize{}3121} & {\scriptsize{}2990}\tabularnewline
\hline 
{\scriptsize{}17} & {\scriptsize{}885} & {\scriptsize{}890} & {\scriptsize{}911} & {\scriptsize{}900} &  & {\scriptsize{}37} & {\scriptsize{}3076} & {\scriptsize{}2992} & {\scriptsize{}3202} & {\scriptsize{}3080}\tabularnewline
\hline 
{\scriptsize{}18} & {\scriptsize{}963} & {\scriptsize{}959} & {\scriptsize{}962} & {\scriptsize{}950} &  & {\scriptsize{}38} & {\scriptsize{}3437} & {\scriptsize{}3434} & {\scriptsize{}3588} & {\scriptsize{}3470}\tabularnewline
\hline 
{\scriptsize{}19} &  & {\scriptsize{}1005} & {\scriptsize{}1018} & {\scriptsize{}1000} &  & {\scriptsize{}39} & {\scriptsize{}3484} & {\scriptsize{}3485} & {\scriptsize{}3632} & {\scriptsize{}3530}\tabularnewline
{\scriptsize{}20} & {\scriptsize{}1031} & {\scriptsize{}1046} & {\scriptsize{}1060} & {\scriptsize{}1040} &  & {\scriptsize{}MAE} &  &  & {\scriptsize{}48(38)} & {\scriptsize{}17 (21)}\tabularnewline
\hline 
\end{tabular}{\scriptsize\par}
\end{table}
 Table \ref{tab:timine_fundamentals} reports the vibrational frequencies
at different level of calculation and compares them with the experimental
ones.. In the Table, DC SCIVR values are compared with the experimental
energies of gas phase isolated and Ar-tagged thymine. We also report
modes 1-3 even if no experimental data are available at the best of
our knowledge. From the Table, we observe that the MAE of DC SCIVR
estimates is equal to 17 and 21 wavenumbers in comparison with gas
phase and Ar-tagged levels respectively, which is less than half the
harmonic estimates. These values are in line once again with the typical
semiclassical accuracy and are comparable with what was found from
previous pyrimidine bases.

\begin{figure}[H]
\centering{}\includegraphics[scale=0.5]{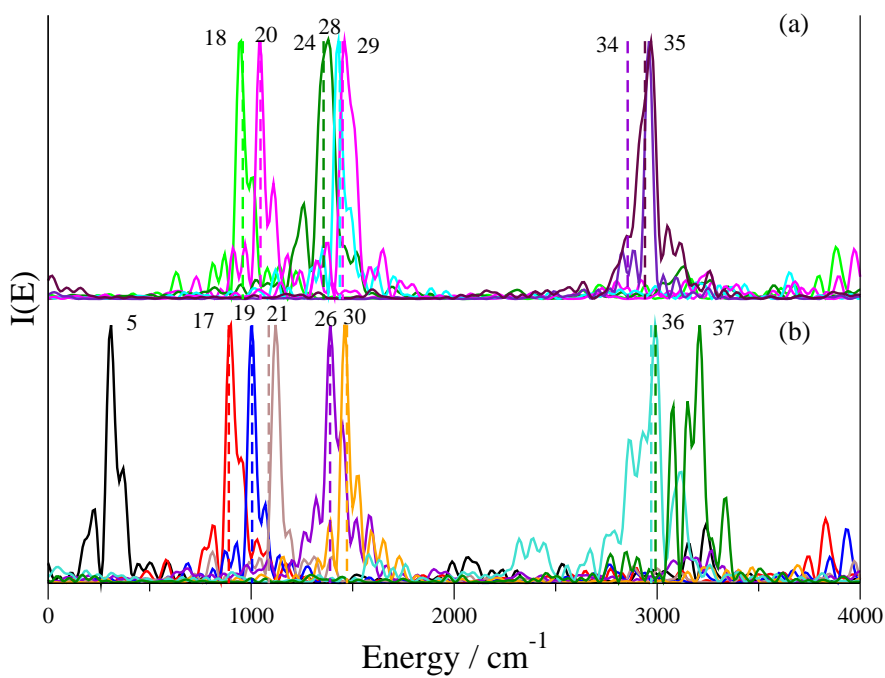}\caption{\label{fig:timine_14D}Panel (a) and (b): The same as in Fig. \ref{fig:Uracil_15D_spectra}
but for the 18-dimension subspace of thymine. Labels of the energy
levels follow the nomenclature reported in Table \ref{tab:timine_fundamentals}.}
\end{figure}
Figure \ref{fig:timine_14D} shows the computed DC SCIVR 18 dimension
spectra and one can notes how the thymine fingerprint region around
1000 wavenumbers is very crowded. However, by using Eq. (\ref{eq:Multiple_Coherent_Ref_State}),
we are able to selectively report each peak, as showed in Fig. \ref{fig:timine_14D}.
In this way it is possible to resolve and attribute peaks which would
be otherwise hardly distinguishable, since they are within few wavenumbers
in energy, as in the case of modes 18-20 and 28-30.

\subsection{Adenine}

\begin{figure}[H]
\centering{}\includegraphics[scale=0.5]{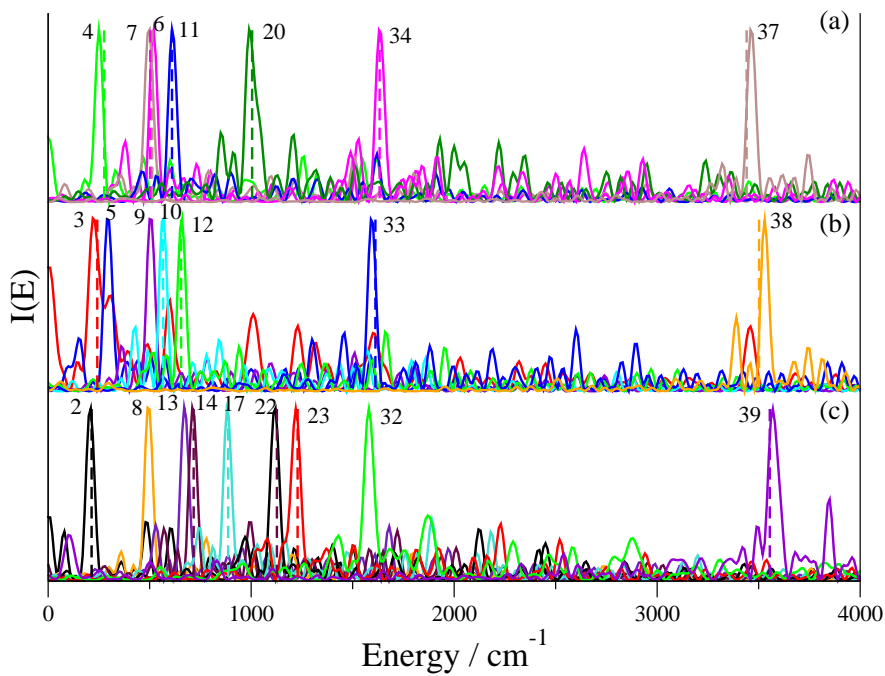}\caption{\label{fig:adenine_14D}Panel (a), (b) and (c): The same as in Fig.
\ref{fig:Uracil_15D_spectra} but for the 23-dimensional subspace
of adenine. Frequency labels follow the nomenclature reported in Table
\ref{tab:timine_fundamentals}.}
\end{figure}

Now we move our attention to the remaining nucleobase, adenine. Adenine
is a nucleobase made of a purine ring, composed of a pyrimidine condensed
with an imidazole. The molecular structure at equilibrium configuration
is reported in panel (e) of Fig. \ref{fig:structure_nucleobases},
showing 15 atoms and resulting into 39 vibrational degrees of freedom,
the same as thymine. In this study we focus our attention only on
the global minimum structure, since previous works suggested that
the other tautomers of the molecule are located around 30 kJ/mol above
the global minimum, an amount of energy difference that is close to
that of thymine and uracil, rather than cytosine. 
\begin{table}[H]
\caption{\label{tab:adenine_fundamentals}Vibrational frequencies for adenine.
First column reports the label of the excitation. Columns two and
three are for gas phase spectra of isolated\citep{colarusso_bernath_IRnucleobasis_1997}
and Ar-tagged adenine.\citep{nowak_Leszczynski_adenineIRartagged_1996}
Column four reports the Perturbation Theory (PT2) energy levels.\citep{biczysko_barone_adenine_2009}
Column five report the harmonic estimate at the level of B3LYP/aug-cc-pvdz
and the last column our computed DC SCIVR energy levels at the same
level of theory. MAE is reported on the last row, where the value
in parenthesis refers to the experiment in Ar matrix. Values are reported
in cm$^{\text{-1}}$.}

\centering{}{\scriptsize{}}%
\begin{tabular}{ccccccccccccc}
{\scriptsize{}mode} & {\scriptsize{}Gas} & {\scriptsize{}Gas/Ar} & {\scriptsize{}PT2} & {\scriptsize{}Harmonic} & {\scriptsize{}DC SCIVR} &  & {\scriptsize{}mode} & {\scriptsize{}Gas} & {\scriptsize{}Gas/Ar} & {\scriptsize{}PT2} & {\scriptsize{}Harmonic} & {\scriptsize{}DC SCIVR}\tabularnewline
\hline 
{\scriptsize{}1} & {\scriptsize{}162} &  & {\scriptsize{}139} & {\scriptsize{}168} & {\scriptsize{}160} &  & {\scriptsize{}21} & {\scriptsize{}1065} & {\scriptsize{}1061} & {\scriptsize{}1054} & {\scriptsize{}1076} & {\scriptsize{}1050}\tabularnewline
\hline 
{\scriptsize{}2} &  & {\scriptsize{}214} & {\scriptsize{}181} & {\scriptsize{}210} & {\scriptsize{}210} &  & {\scriptsize{}22} & {\scriptsize{}1126} & {\scriptsize{}1127} & {\scriptsize{}1125} & {\scriptsize{}1141} & {\scriptsize{}1120}\tabularnewline
\hline 
{\scriptsize{}3} & {\scriptsize{}244} & {\scriptsize{}242} & {\scriptsize{}209} & {\scriptsize{}267} & {\scriptsize{}220} &  & {\scriptsize{}23} & {\scriptsize{}1234} & {\scriptsize{}1229} & {\scriptsize{}1230} & {\scriptsize{}1242} & {\scriptsize{}1220}\tabularnewline
\hline 
{\scriptsize{}4} & {\scriptsize{}270} & {\scriptsize{}276} & {\scriptsize{}276} & {\scriptsize{}279} & {\scriptsize{}250} &  & {\scriptsize{}24} &  & {\scriptsize{}1246} & {\scriptsize{}1243} & {\scriptsize{}1263} & {\scriptsize{}1250}\tabularnewline
\hline 
{\scriptsize{}5} &  &  & {\scriptsize{}299} & {\scriptsize{}302} & {\scriptsize{}290} &  & {\scriptsize{}25} & {\scriptsize{}1280} & {\scriptsize{}1290} & {\scriptsize{}1291} & {\scriptsize{}1332} & {\scriptsize{}1300}\tabularnewline
\hline 
{\scriptsize{}6} & {\scriptsize{}506} & {\scriptsize{}503} & {\scriptsize{}491} & {\scriptsize{}512} & {\scriptsize{}520} &  & {\scriptsize{}26} & {\scriptsize{}1326} & {\scriptsize{}1328} & {\scriptsize{}1325} & {\scriptsize{}1356} & {\scriptsize{}1330}\tabularnewline
\hline 
{\scriptsize{}7} & {\scriptsize{}515} & {\scriptsize{}513} & {\scriptsize{}516} & {\scriptsize{}529} & {\scriptsize{}500} &  & {\scriptsize{}27} & {\scriptsize{}1346} & {\scriptsize{}1345} & {\scriptsize{}1338} & {\scriptsize{}1365} & {\scriptsize{}1330}\tabularnewline
\hline 
{\scriptsize{}8} &  &  & {\scriptsize{}518} & {\scriptsize{}531} & {\scriptsize{}490} &  & {\scriptsize{}28} &  & {\scriptsize{}1389} & {\scriptsize{}1376} & {\scriptsize{}1414} & {\scriptsize{}1390}\tabularnewline
\hline 
{\scriptsize{}9} &  &  & {\scriptsize{}529} & {\scriptsize{}542} & {\scriptsize{}500} &  & {\scriptsize{}29} & {\scriptsize{}1415} & {\scriptsize{}1419} & {\scriptsize{}1406} & {\scriptsize{}1432} & {\scriptsize{}1400}\tabularnewline
\hline 
{\scriptsize{}10} & {\scriptsize{}563} & {\scriptsize{}566} & {\scriptsize{}570} & {\scriptsize{}583} & {\scriptsize{}570} &  & {\scriptsize{}30} & {\scriptsize{}1468} & {\scriptsize{}1474} & {\scriptsize{}1466} & {\scriptsize{}1498} & {\scriptsize{}1485}\tabularnewline
\hline 
{\scriptsize{}11} & {\scriptsize{}600} & {\scriptsize{}610} & {\scriptsize{}610} & {\scriptsize{}617} & {\scriptsize{}610} &  & {\scriptsize{}31} &  & {\scriptsize{}1482} & {\scriptsize{}1481} & {\scriptsize{}1512} & {\scriptsize{}1500}\tabularnewline
\hline 
{\scriptsize{}12} & {\scriptsize{}650} & {\scriptsize{}655} & {\scriptsize{}655} & {\scriptsize{}668} & {\scriptsize{}660} &  & {\scriptsize{}32} &  &  & {\scriptsize{}1577} & {\scriptsize{}1607} & {\scriptsize{}1580}\tabularnewline
\hline 
{\scriptsize{}13} &  &  & {\scriptsize{}677} & {\scriptsize{}688} & {\scriptsize{}670} &  & {\scriptsize{}33} &  & {\scriptsize{}1612} & {\scriptsize{}1591} & {\scriptsize{}1635} & {\scriptsize{}1590}\tabularnewline
\hline 
{\scriptsize{}14} &  & {\scriptsize{}717} & {\scriptsize{}717} & {\scriptsize{}723} & {\scriptsize{}710} &  & {\scriptsize{}34} & {\scriptsize{}1625} & {\scriptsize{}1633} & {\scriptsize{}1616} & {\scriptsize{}1660} & {\scriptsize{}1630}\tabularnewline
\hline 
{\scriptsize{}15} & {\scriptsize{}801} & {\scriptsize{}802} & {\scriptsize{}811} & {\scriptsize{}815} & {\scriptsize{}800} &  & {\scriptsize{}35} &  & {\scriptsize{}3041} & {\scriptsize{}3049} & {\scriptsize{}3172} & {\scriptsize{}3070}\tabularnewline
\hline 
{\scriptsize{}16} & {\scriptsize{}847} & {\scriptsize{}848} & {\scriptsize{}846} & {\scriptsize{}858} & {\scriptsize{}840} &  & {\scriptsize{}36} & {\scriptsize{}3061} & {\scriptsize{}3057} & {\scriptsize{}3102} & {\scriptsize{}3245} & {\scriptsize{}3160}\tabularnewline
\hline 
{\scriptsize{}17} &  & {\scriptsize{}887} & {\scriptsize{}885} & {\scriptsize{}896} & {\scriptsize{}880} &  & {\scriptsize{}37} & {\scriptsize{}3434} & {\scriptsize{}3441} & {\scriptsize{}3432} & {\scriptsize{}3588} & {\scriptsize{}3460}\tabularnewline
\hline 
{\scriptsize{}18} & {\scriptsize{}926} & {\scriptsize{}927} & {\scriptsize{}931} & {\scriptsize{}942} & {\scriptsize{}930} &  & {\scriptsize{}38} & {\scriptsize{}3501} & {\scriptsize{}3502} & {\scriptsize{}3497} & {\scriptsize{}3641} & {\scriptsize{}3530}\tabularnewline
\hline 
{\scriptsize{}19} & {\scriptsize{}957} & {\scriptsize{}958} & {\scriptsize{}969} & {\scriptsize{}979} & {\scriptsize{}960} &  & {\scriptsize{}39} & {\scriptsize{}3552} & {\scriptsize{}3555} & {\scriptsize{}3539} & {\scriptsize{}3727} & {\scriptsize{}3570}\tabularnewline
\hline 
{\scriptsize{}20} &  & {\scriptsize{}1005} & {\scriptsize{}1018} & {\scriptsize{}1015} & {\scriptsize{}990} &  & {\scriptsize{}MAE} &  &  & {\scriptsize{}10 (9)} & {\scriptsize{}42 (38)} & {\scriptsize{}16 (14)}\tabularnewline
\hline 
\end{tabular}{\scriptsize\par}
\end{table}
 The 39 vibrational degree of freedom space is separated into one
23-dimensional, two bi-dimensional and all other are mono-dimensional
subspaces, by employing a threshold parameter for the Hessian method
equal to $\varepsilon=9\cdot10^{-7}$. Figure \ref{fig:adenine_14D}
reports the spectrum of the highest dimensional subspace for this
molecule.
 Once again, our reference state choice according to Eq. (\ref{eq:Multiple_Coherent_Ref_State})
helps us to resolve different peaks which are very close in energy,
like modes 37, 38 and 39 or 32, 33 and 34. Fig. \ref{fig:adenine_14D}
presents several overtone peaks, in comparison with previous ones.
We believe this is due to the presence of the $\text{NH}_{2}$ hindered
rotation, which gets easily vibrational excited during the dynamics.
Nevertheless, the full width at half maximum (FWHM), which denotes
the peak definition, is quite small. For a more detailed comparison
with experiments, Table \ref{tab:adenine_fundamentals} reports the
computed energy levels and shows the comparison with available theoretical
and experimental results, measured in isolated gas phase and Ar matrix.
The average deviation of the DC SCIVR results from both experiments
is quite small, only 16 and 14 wavenumbers, almost three times more
accurate than harmonic estimates. This accuracy is comparable with
previous system investigations and with PT2 estimates. Overall, the
DC SCIVR spectrum of adenine is a further proof of the invariance
of accuracy of our approach with the increase in nucleobase dimensionality.
We believe that the investigation of this purine nucleobase strengthens
previous considerations about pyrimidine-based nucleobases, in terms
of spectroscopic quality and accuracy of the energy levels, when compared
with the experiments.

\section{Summary and Conclusions\label{sec:Conclusions}}

In this paper we have presented a semiclassical investigation of the
vibrational features of uracil, cytosine, thymine, and adenine nucleobases.
The investigation on uracil has shown that MC SCIVR energy levels
are on average around 20 wavenumbers away from experimental levels,
a typical value for semiclassical spectroscopic calculations. Then,
the DC SCIVR method leads to values very close to the full-dimensional
MC SCIVR ones, proving its reliability for the calculation of similar
systems. Moving to higher dimensional nucleobases, DC SCIVR energy
levels of cytosine retain their typical accuracy with respect to the
experimental results. Despite the presence of more than one comparable
minimum, the method still reproduces the spectra of different isomers
retaining the standard accuracy of semiclassical simulations. Then,
we focus on the last pyrimidine molecule, thymine, which is the highest
dimensional nucleobase of this type. We have found that DC SCIVR retains
its accuracy despite the increased dimensionality. Similar considerations
hold for the adenine case, where the MAE is around 15 wavenumbers
and comparable with PT2 estimates. Overall, we always find the accuracy
of the DC SCIVR method to be comparable to other state of the art
theoretical spectroscopy methods.

These outcomes are promising for a future exploitation of the method.
Since the accuracy is seemingly insensitive to the increase in the
molecule dimensionality, we will exploit DC SCIVR also to study more
complex systems, like nucleotides and nucleobase pairs. This will
possibly pave the way toward the assessment of important structural
features of nucleoacids that lead to the formation of secondary and
tertiary structures.

\section*{Acknowledgments}

Authors thank dr. Riccardo Conte for useful discussions. Authors acknowledge
support from the European Research Council (ERC) under the European
Union\textquoteright s Horizon 2020 research and innovation programme
(grant agreement No {[}647107{]} -- SEMICOMPLEX -- ERC-2014-CoG).
Authors acknowledge also CINECA (Italian Supercomputing Center) for
providing high perfomance computational resources under the IscraB
grant (QUASP). All authors thank Università degli Studi di Milano
for further computational time at CINECA.

\newpage

\bibliographystyle{aipnum4-1}
\bibliography{SEMICOMPLEX}

\end{document}